\documentclass[11pt]{article}
\usepackage{moriond,here,epsfig,amssymb}

\def\Journal#1#2#3#4{{#1} {\bf #2}, #3 (#4)}

\def\NIMA{{\em Nucl. Instrum. Methods} A}

\def\PLB{{\em Phys. Lett.}  B}

\def\EPJC{{\em Eur. Phys. J.} C}

\def\ra{$\rightarrow$ }

\def\epem{$\mathrm{e^+e^-}$}
\def\ll{$\mathrm{l^+l^-}$}
\def\mumu{$\mathrm{\mu^+\mu^-}$}
\def\tautau{$\mathrm{\tau^+\tau^-}$}
\def\nunu{$\mathrm{\nu^+\nu^-}$}
\def\tp{$\mathrm{\tau^+}$}
\def\tm{$\mathrm{\tau^-}$}
\def\bb{$\mathrm{b\bar{b}}$}
\def\qq{$\mathrm{q\bar{q}}$}
\def\csbar{$\mathrm{c\bar{s}}$}
\def\cbars{$\mathrm{\bar{c}s}$}
\def\tpnu{$\mathrm{\tau^+\nu}$}
\def\tmnu{$\mathrm{\tau^-\bar{\nu}}$}

\def\Hpm{$\mathrm{H^{\pm}}$}
\def\mHpm{$\mathrm{m_{H^{\pm}}}$}

\def\h{$\mathrm{h^0}$}
\def\mh{$\mathrm{m_h}$}
\def\Zo{$\mathrm{Z^0}$}

\def\So{$\mathrm{S^0}$}
\def\Ho{$\mathrm{H^0}$}
\def\A{$\mathrm{A^0}$}
\def\Hp{$\mathrm{H^+}$}
\def\Hm{$\mathrm{H^-}$}

\def\Hone{$\mathrm{H_1}$}
\def\Htwo{$\mathrm{H_2}$}
\def\Hthree{$\mathrm{H_3}$}
\def\mA{$\mathrm{m_A}$}
\def\mH{$\mathrm{m_H}$}
\def\mS{$\mathrm{m_S}$}
\def\sba{$\mathrm{\sin^2(\beta-\alpha)}$}
\def\tanb{$\mathrm{\tan\beta}$}
\def\sqrts{$\mathrm{\sqrt{s}}$}
\def\cba{$\mathrm{\cos^2(\beta-\alpha)}$}
\def\gevc{GeV/$c^2$}

\def\be{\begin{equation}}
\def\ee{\end{equation}}
\def\bea{\begin{eqnarray}}
\def\eea{\end{eqnarray}}

\begin{document}
\vspace*{4cm}
\title{Higgs Boson Searches at LEP}

\author{ Pamela Ferrari }

\address{CERN, 1211 Geneve 23, Switzerland}

\maketitle\abstracts{
The results of the Higgs boson searches performed by the four LEP experiments
at centre-of-mass energies between 189 GeV and 209 GeV corresponding to an
integrated luminosity of 2461 pb$^{-1}$ are presented here. Searches 
have been performed for Higgs in the Standard Model (SM), in 2 Higgs Doublet Models
(2HDM's), for doubly charged, fermiophobic and invisible Higgs as well
as a decay mode independent search. Most of the results of the four
experiments have been combined by the LEP Higgs working group. }

\section{Introduction}
\label{sec:intro}

We present combined results from the
ALEPH, DELPHI, L3 and OPAL Collaborations on searches for the 
SM Higgs boson, for the neutral Higgs bosons \h\ and \A\ in the Minimal 
Supersymmetric Standard Model (MSSM),
for charged Higgs bosons, for ``fermiophobic''
Higgs decaying into a pair of photons.
Individual experiments have prepared analyses for future combinations: 
 searches for neutral Higgs bosons \h\ and \A\ in different types of 2HDM's, 
for charged Higgs decaying into $\mathrm {A^0, W^{\pm*}}$, for doubly charged 
Higgs as well as a decay mode independent search are presented.
The analyses are based on 2461pb$^{-1}$ of \epem collisions at LEP2, for
 centre-of-mass energies between 189 and 209 GeV.
Each experiment has generated
Monte Carlo event samples for the Higgs signal and the various 
background processes, typically,
at 189, 192, 196, 200, 202, 204, 206, 208 and 210~GeV energies. 
Cross-sections, branching ratios, 
distributions of the reconstructed mass and other 
discriminating variables relevant to the combination have been interpolated 
to energies which 
correspond to the data sets. In this procedure special care has been taken 
to the regions of kinematic cutoff where the signal and
background distributions vary rapidly. It has been established that the 
interpolation procedures do not add significantly
to the final systematic errors.\\
The main sources of systematic error affecting the signal and 
background rate predictions are included taking into account correlations between search channels, LEP energies and individual experiments. 
This is done using an
extension of the method of Cousins and Highland~\cite{cousins-highland} where 
the confidence levels are the averages of a large ensemble of Monte Carlo 
experiments, each one with a different choice of
signal and background, varied within the errors.

\section{Search for the SM Higgs boson}

At  LEP  the  SM Higgs  boson  is  expected  to  be  produced mainly  via  the
{\it Higgs--strahlung}  process  \epem\ra~\Ho\Zo,  while  contributions  from  the
 WW\ra~H  fusion channel,  \epem\ra~\Ho$\nu_e\bar{\nu_e}$,  are typically  below
 10\%.  The  searches performed by  the four LEP collaborations  encompass the
 usual   \Ho\Zo\   final   state   topologies,  commonly   called   `four-jet'
 (\Ho\Zo\ra\bb\qq),  `missing  energy'  (\bb\nunu),  `leptonic'  (\bb\epem\  and
 \bb\mumu) and `tau' channels (\bb\tp\tm\ and \tp\tm\qq).  The searches in the
 missing  energy  channel are  optimised  for  {\it Higgs--strahlung},  but are  also
 sensitive to the WW\ra~H fusion process. The analyses are based on 2461
 pb$^{-1}$ of data taken at centre of mass energies 189 $\le$ \sqrts $\le$ 209 GeV.
The results of the final combination of the LEP results on SM Higgs searches 
is presented here, more details can be found in~\cite{lephiggs}. 
The  combined   LEP  data   are  used  to   test  two  hypotheses:   the  {\it
background-only}  (``$b$") hypothesis,  which  assumes no  Higgs  boson to  be
present  in the  mass range  investigated, and  the  {\it signal~+~background}
(``$s+b$") hypothesis, where Higgs bosons are assumed to be produced according
to  the  model  under   consideration.   A  global  {\it  test-statistic}  $Q$
\cite{lephiggs}   is  constructed   which  allows   the   experimental  result
$Q_{observed}$  to   be  classified   between  the  $b$-like   and  $s+b$-like
situations.  
The probability density functions of $Q$(\mH) are integrated from the observed
value to plus or minus infinity to
form the confidence levels $CL_{s+b}$ and $CL_b$, which express 
the probabilities that the outcome of an experiment is more $s+b$-like or 
less $b$-like, respectively, than the outcome represented by the set of 
selected events. To prevent to extend the limit beyond the range of 
sensitivity the ratio $CL_s$=$CL_{s+b}$/$CL_b$ is used 
to set mass exclusion limits. 
\begin{figure}[H]
\begin{center}
\epsfig{figure=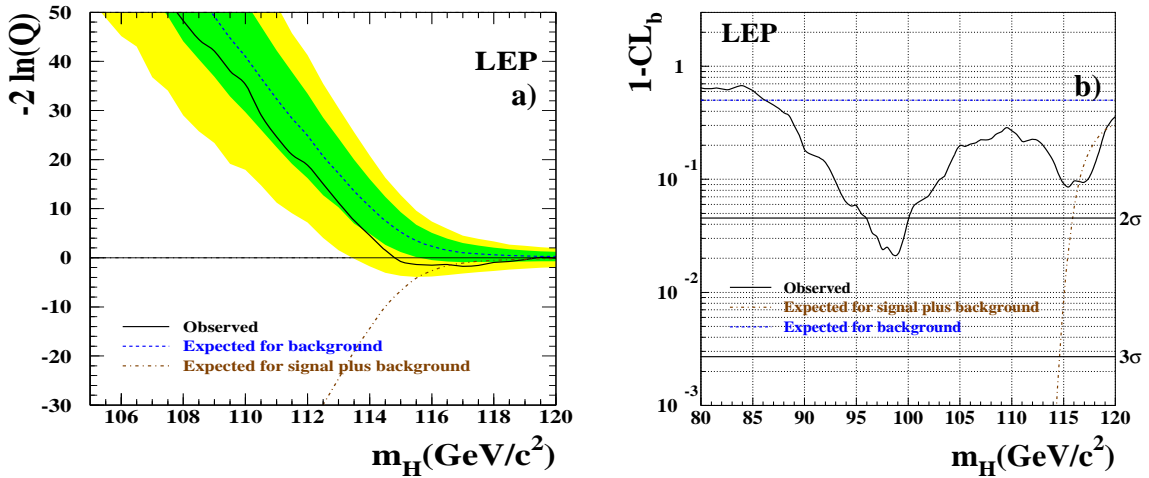,width=1\textwidth}
\end{center}
\caption{a) The test-statistic as a function of
\mH. Dashed line: expectation and $1\sigma$ and 
$2\sigma$ probability bands around it in the background only hypothesis;
solid curve: observed results; dotted line: median result expected 
in presence of signal where \mH=115 \gevc.
b) $1-CL_b$ as a function of \mH. 
Straight line at 50\%: median result in the absence of a
signal. Solid curve: observed result. Dashed curve:
median result expected for a signal when tested at the ``true" mass.
\label{fig:2lnQ}}
\end{figure}

Figure~\ref{fig:2lnQ}.a shows the test-statistic -2ln$Q$
versus the test mass \mH\ for the combined LEP data: 
there is a broad minimum in the observed curve starting at about 115
GeV/$c^2$. The expectation for
signal plus background hypothesis crosses the observed curve 
close to this value, indicating that a Higgs boson with such mass is
more favoured than the ``$b$" hypothesis, albeit at low significance. 
Studies of the contributions from individual experiments and final-state
topologies have shown that this signal-like behaviour mainly originates
from the ALEPH data in the four-jet channel. 
Figure~\ref{fig:2lnQ}.b shows the confidence level $1-CL_b$ for test masses
in the range 80-120 GeV/$c^2$. In the region \mH$\sim$ 98 \gevc\ the deviation
from the background only hypothesis corresponds to 2.3 standard deviations.
In the region of \mH\ above 115 \gevc\ it corresponds to 1.7 standard 
deviations. This deviation, though of low significance is compatible
with a SM Higgs boson signal, while being also in agreement with the 
background hypothesis. 
A 95\% confidence level lower limit on the Higgs
mass may be set by identifying the mass region where $CL_s < 0.05$:
the median limit expected in the absence of a signal is 115.3~\gevc\
and the observed limit is 114.4~\gevc. 

\section{Bounds for the Higgs boson coupling}

The combined LEP data are also used to set 95\% confidence level upper bounds
on the HZZ coupling in non-standard models~\cite{lephiggs}.
The limits are expressed in terms of
$\xi ^2$ defined as the ratio
of the non-standard HZZ coupling and the same coupling in the SM.
In deriving these limits LEP1 data collected at the Z resonance  have been 
combined with LEP2 data between 161 and 209 GeV.
This limit is valid only in the hypothesis that the Higgs boson
has SM branching ratios, but extensions of the SM could easily predict 
suppressed couplings to b-quarks, therefore flavour independent
searches have been performed by the four LEP collaborations, analysing
the four jet (\qq\qq), missing energy (\qq\nunu) and leptonic (\qq\ll)
topologies. Data collected for 189 $\le$ \sqrts $\le$ 209 GeV have been analysed
without finding any evidence of the presence of a signal~\cite{flavind}.
The coupling limits obtained by the individual experiments
are not far from the ones provided by the usual SM searches, searching for final
states containing b-quarks.
The observed (median expected) limits on the Higgs mass assuming
SM production cross-sections are 110.6 (110.5) GeV, 110.6 (108.0)
GeV, 108.7 (110.3) GeV and 109.2 (108.0) GeV for ALEPH, DELPHI, L3 and OPAL,
respectively. A LEP combination of the results is expected soon.

\section{Two Higgs Doublet Models}

It is important to study extensions of the SM  
containing more than one physical Higgs boson in the spectrum:
in particular Two Higgs Doublet Models (2HDMs) which are attractive 
since they add new phenomena with the fewest new parameters and since
the MSSM is a special case of 2HDM in which the addition of supersymmetry
adds new particles and constrains the model itself.
In the context of general 2HDMs
the Higgs sector comprises five physical Higgs bosons: 
two neutral CP-even scalars, \h\ and \Ho\ (with  \mh\ $<$ \mH), one
CP-odd scalar, \A, and two charged scalars, \Hpm. 
At the centre-of-mass energies accessed by
LEP, the \h\ and \A\  
bosons are expected to be produced predominantly via two processes: 
the {\it{Higgs--strahlung}}
process, \epem\ra\h\Zo, 
and the {\it{pair--production}} process, \epem\ra\h\A.
The cross-sections for these two processes,
$\sigma_{\mathrm{hZ}}$ and $\sigma_{\mathrm{hA}}$,
are related at tree-level 
to the SM cross-sections by the following relations: 
\be
\sigma_{\mathrm{hZ}}=\sin^2(\beta -\alpha)~\sigma^{\mathrm{SM}}_{\mathrm{HZ}},~
\sigma_{\mathrm{hA}}=
\cos^2(\beta-\alpha)~\bar{\lambda}~\sigma^{\mathrm{SM}}_{\mathrm{HZ}},
\label{equation:xsec_ah}
\ee

\vspace{-0.15cm}\hspace{-0.55cm}where $\sigma^{\mathrm{SM}}_{\mathrm{HZ}}$ is
the 
{\it Higgs--strahlung} cross-section for the SM process \epem\ra\Ho\Zo, and
$\bar{\lambda}$ is a phase--space factor.\\
Within 2HDMs the choice of the couplings between the Higgs bosons and
the fermions determines the type of the model considered. In the Type II 
model the first Higgs doublet ($\phi_1$)
couples only to down--type fermions and the 
second Higgs doublet ($\phi_2$) couples only to up--type 
fermions. In the Type I model the 
quarks and leptons do not couple to the first Higgs 
doublet ($\phi_1$), but couple to the second Higgs 
doublet ($\phi_2$).
\begin{figure}[H]
\epsfig{figure=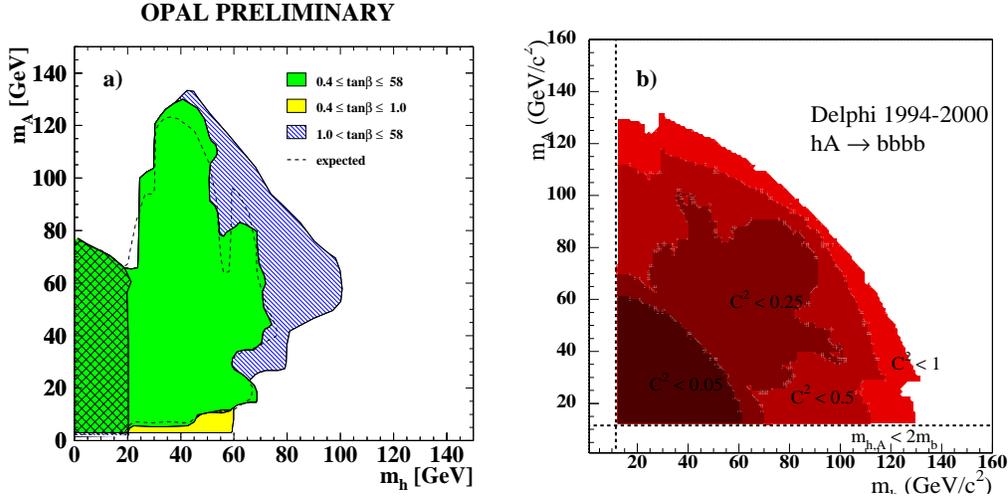,height=7.0cm}
\caption { Figure~\ref{fig:2hdm}: a) Excluded ($\mathrm{m_A}$,$\mathrm{m_h}$)
region independent of $\alpha$ ( solid line ) and expected exclusion 
( dashed line). Different $\mathrm{tan\beta}$ intervals have been considered.
The cross-hatched region  is excluded using $\Gamma_Z$ constrains only.
b) Excluded $\mathrm {c^2}$ contours in the (\mA,\mh) plane assuming that \h\
and \A\ decay to \bb\ in 100\% of the cases.
\label{fig:2hdm}}
\end{figure}
The Higgs sector in the MSSM is a 2HDM(II).\\
In a 2HDM the production cross-sections and Higgs boson decay branching ratios
are predicted for a given set of model parameters.
The coefficients \sba\ and \cba\ which appear
in Eq.~(\ref{equation:xsec_ah})
determine the production cross-sections.  The decay branching ratios to
the various final states 
are also determined by $\alpha$ and $\beta$.
In the 2HDM(II) the tree-level couplings of the \h\ and \A\ bosons to the up-- and down--type 
quarks relative to the canonical SM values are:
\be
{\mathrm{h^0}} {\mathrm{c}} \overline{{\mathrm{c}}} = {\cos \alpha}/
{\sin\beta},~~{\mathrm{h^0}} 
{\mathrm{b}} \overline{{\mathrm{b}}} = -~{\sin \alpha}/{\cos\beta},
~~{\mathrm{A^0}} {\mathrm{c}} \overline{{\mathrm{c}}} = \cot \beta,~~
{\mathrm{A^0}} {\mathrm{b}} \overline{{\mathrm{b}}} = \tan \beta.
\label{equation:br2hdm}
\ee
From these equations it is easy to see that in wide portions
of the parameter space the couplings to b-quarks are not the dominant ones.\\
The OPAL collaboration has performed a 
detailed scan over broad ranges of these parameters~\cite{2hdm}:
$1\le$ \mh $\le 100$ GeV, $5 \le$ \mA $\le 2$ TeV, $0.4 \le $ \tanb $\le 58.0$
and $\alpha = 0, -\pi/8, -\pi/4, -3\pi/8$ and $-\pi/2$.
All available neutral Higgs searches for centre-of-mass energies 
189 $\le$ \sqrts $\le$ 209 GeV have been combined with LEP1 data collected at the
Z peak. Both analyses making use of b-tagging and flavour independent searches
have been used. The most general exclusion obtained by OPAL is shown in
Figure~\ref{fig:2hdm}.a.
The DELPHI collaboration has also produced limits in the 
(\mA,\mh) plane on $\mathrm {c^2}$, defined as the ratio of the 2HDM production 
cross-section for the process \epem\ra\h\A\ and the maximal production
cross-section~\cite{2hdmdelphi}. These limits are obtained with the assumption
of 100\% decays into the following specific final states:
\h\A\ra\bb\bb, \h\A\ra\A\A\A\ra\bb\bb\bb, \h\Zo\ra\A\A\Zo\ra\bb\bb\qq, the excluded
contours for the \h\A\ra\bb\bb\ case are shown in Figure~\ref{fig:2hdm}.b.

\subsection{MSSM Higgs searches}
The MSSM is a 2HDM(II) in which supersymmetry adds new particles and
constrains the model. The production cross-sections and decay branching ratios
depend not only on the masses but also on the values of $\alpha$ and $\beta$
as in equations ~\ref{equation:xsec_ah} and ~\ref{equation:br2hdm}.
In most of the parameter space the decays of \h\ and \A\ into \bb\ and \tautau\
dominate, therefore the searches for MSSM signatures concentrate
on these final states, but for the values of parameters for which
these decays are suppressed also flavour-independent searches
are taken into account. 
The individual searches of the four LEP collaborations for the processes 
\epem\ra~\h\Zo\ and \epem\ra~\h\A\ 
 which include the data taken at 88 $\le$ \sqrts $\le$ 209~GeV, have been combined as
described in~\cite{mssm}. No evidence for the presence of a signal has been
discovered. Exclusion limits have been obtained in 
three `benchmark' MSSM parameter 
scans.
The first benchmark corresponds 
to {\it no-mixing} in the scalar-top sector; a second to large mixing 
and the parameters tuned to maximise the parameter space along \mh\ 
({\it \mh-max} hereafter); a third scan ({\it large-$\mu$} hereafter)
is designed to highlight choices of MSSM parameters for which the \h\
does not decay into \bb\ due to large loop corrections.
\begin{figure}[H]
\begin{center}
\epsfig{figure=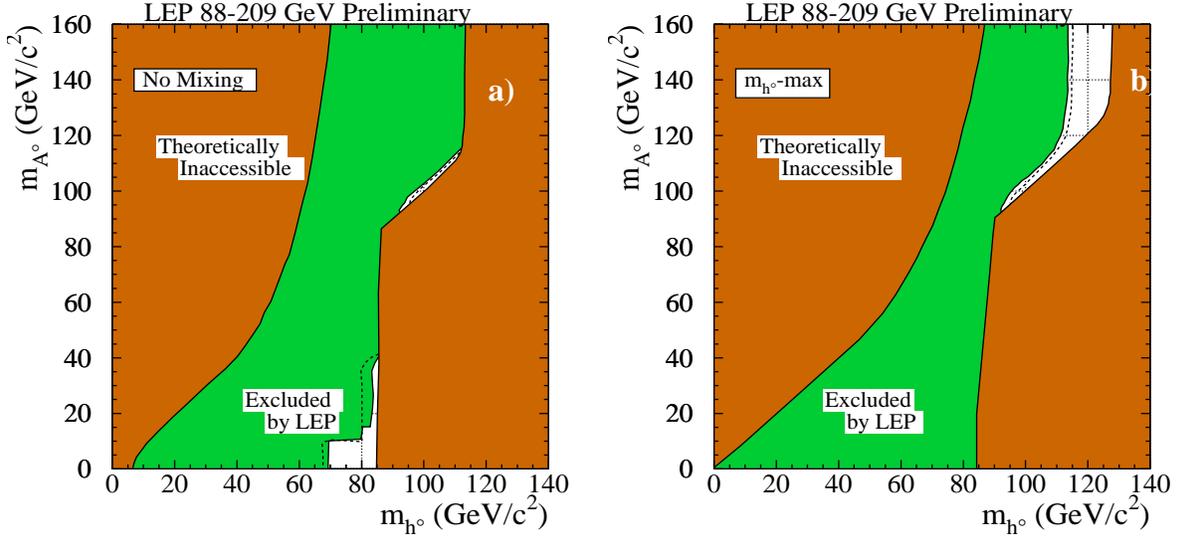,width=1\textwidth}
\end{center}
\caption{The MSSM exclusion (a) for the no-mixing scenario and (b) for the 
 {\it \mh-max} scenario. The figures show the excluded and the
 theoretically not allowed regions as a function of the Higgs boson mass. 
\label{fig:mssm}}
\end{figure}
Figures~\ref{fig:mssm}.a) and~\ref{fig:mssm}.b)  show the excluded regions in 
the {\it no-mixing} and in the {\it \mh-max} scenarios, respectively.
The {\it no-mixing} scenario is almost completely excluded except for
the small region for low values of \mA\ and \mh $\ge$ 60 GeV. In the final
forthcoming LEP combination, the unexcluded area will be considerably reduced: 
the region for 2 $\le$ \mA $\le$ 10 \gevc\ is already excluded
by a new search by OPAL for a low mass A in the \epem\ra\h\Zo\ra\A\A\Zo\
process~\cite{lowma}, which is dominant in this mass range.
The  region for \mA $\ge$ 12 and \mh $\ge$ 80 GeV has been excluded by the latest
MSSM scan performed by the DELPHI collaboration~\cite{mssmdelphi}.
The {\it large-$\mu$} scenario has been completely excluded at 95\% CL
by using the flavour independent searches.

\subsection{CP violating MSSM Higgs searches}

The introduction of CP violating phases into the MSSM is theoretically appealing, since 
CP violation is one of the three requirements to be fulfilled 
to generate the cosmic matter/antimatter asymmetry.
CP violating phases in the sector of direct soft supersymmetry
breaking lead to an introduction of CP violation to the MSSM Higgs sector via
first order loop corrections from third generation squarks to the otherwise CP invariant 
Higgs potential. This would allow the MSSM to fulfil the requirements for generating 
the cosmic baryon asymmetry. 
\begin{figure}[H]
\begin{center}
\epsfig{figure=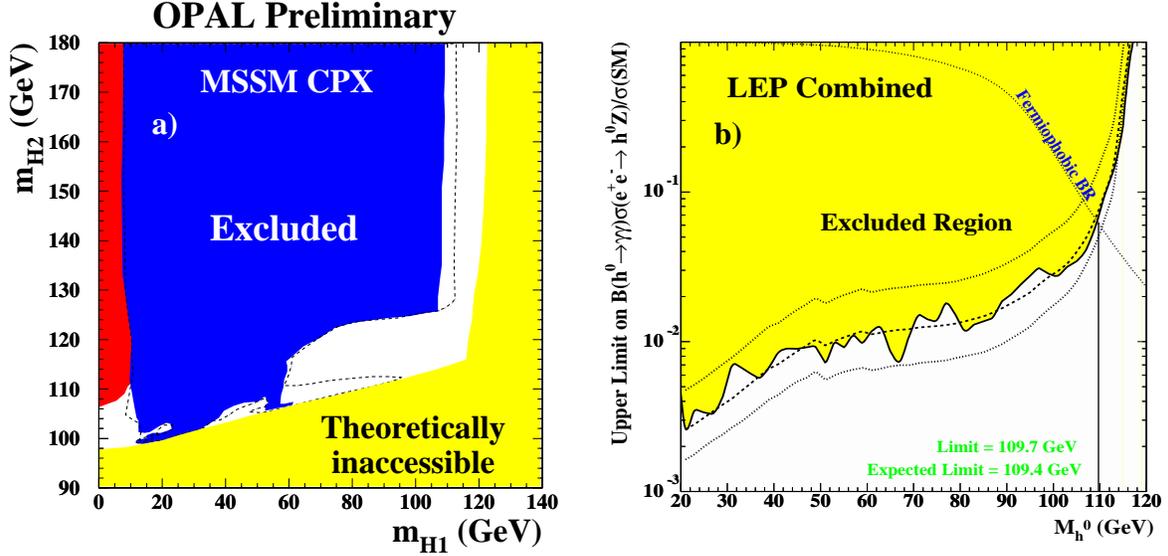,width=\textwidth}
\end{center}
\caption{ a) 95\% CL excluded areas in the CP violating MSSM scenario.
The observed excluded region is shown in the 
($\mathrm{m_{H_1}}$,$\mathrm{m_{H_2}}$).
The expected excluded region is indicated by the dashed line and the
theoretically inaccessible region is light shaded.
b) Combined LEP experimental limits for Higgs bosons decaying into
  di-photons. The 95\% CL confidence level upper limit on the 
B(\h\ra $\gamma\gamma$) x $\sigma$ (\epem \ra\h\Zo)/ $\sigma$(SM) is shown as a
  function of the Higgs mass. Also shown ( dotted line) is the branching
  fraction obtained for the benchmark fermiophobic model. The median
expected limits and the $\pm2\sigma$ confidence level region are denoted by
the dashed curves. The combined limit is indicated by the vertical line.\label{fig:cpvmssm}}
\end{figure}
In a CP violating MSSM scenario, the Higgs production mechanisms proceed as in a CP conserving MSSM. 
As the Higgs mass eigenstates do not have defined CP quantum numbers, the production of all three
mass eigenstates in {\it Higgs-strahlung} is allowed and the coupling of the Higgs
bosons to the \Zo\ is modified. For some choices of the parameters, \Hone, 
decouples completely from the \Zo,
while the production of the second lightest Higgs,\Htwo, also has small or vanishing 
cross-section, allowing for more complex experimental situations and reducing
the accessible parameter regions. 
In large parts of the parameter space the production of both \Hone\ and \Htwo\ in 
{ \it Higgs-strahlung} is possible. The production of the heaviest Higgs state,
\Hthree, has no relevant cross-section at LEP energies in all production
channels for the scenario under study.
The OPAL collaboration has reinterpreted all the searches for neutral Higgs
bosons performed for 88 $\le$ \sqrts $\le$ GeV in this context to constrain the
parameter space of CP violating MSSM scenarios~\cite{CPX}.
The choices of the parameters used for the benchmark scans are fulfilling
the electric dipole moment (EDM) constraints and maximise the CP violating effects.
The results of the scans are shown in Figure~\ref{fig:cpvmssm}

\section{Charged Higgs}
Charged Higgs bosons are predicted by 2HDM's.
The present searches for charged Higgs bosons are placed in the general
context of 2HDM's where the mass is not constrained, since
in  the MSSM at  tree-level the \Hpm\  is 
constrained to be heavier than the W$^{\pm}$ bosons and only for 
extreme choices of the parameters loop corrections can drive the 
mass to lower values. 
At LEP2 energies charged Higgs bosons are expected to be
produced  mainly  through  the  process \epem\ra\Hp\Hm. 
In the 2HDM at tree level the production cross-section is fully determined by
\mHpm;

\subsection{Search for the \Hp\ra\csbar\ and \Hp\ra\tpnu\ decays}

The LEP collaborations have searched for charged Higgs bosons under the
assumption that the two decays \Hp\ra\csbar\ and \Hp\ra\tpnu\ exhaust the \Hp\ decay
width~\cite{charged}; however, the relative branching ratio is not predicted. Thus, the 
searches encompass the following \Hp\Hm\ final states: (\csbar)(\cbars),
(\tpnu)(\tmnu) and the mixed mode (\csbar)(\tmnu)+(\cbars)(\tpnu).
The combined search results are presented as a function of the branching 
ratio B(\Hp\ra\tpnu).
As a preliminary results the L3 collaboration observes
a 2.7$\sigma$ deviation from the 
background behaviour for \mHpm$\simeq$ 68 \gevc, which has not been 
confirmed by the other collaborations. This effect is under investigation.
The 95\% CL observed (expected) lower limits on the charged Higgs 
boson mass for any value of B(\Hp\ra\tpnu) obtained by the other 
LEP experiments are: 79.3 (77.1) \gevc\ for ALEPH, 74.3 (76.4) \gevc\ 
for DELPHI, 75.5 (74.5) \gevc\ for OPAL. The final LEP combination is 
expected soon. \\
\subsection{Search for the \Hp\ra\A$\mathrm{W^{+*}}$ decays}
The OPAL and DELPHI collaborations have also searched for
the process \epem\ra\Hp\Hm\ra\A$\mathrm{W^{*+}}$\A$\mathrm{W^{*-}}$, which is 
relevant at low \tanb\ in 2HDM(II) and dominant for \tanb $\le$ 1 in 2HDM(I),
if kinematically allowed~\cite{aw}.
These searches are restricted to values of \mA\ larger than 12 \gevc, since
the \A\ is assumed to decay into \bb\ final states, while leptonic and hadronic
decays of the the W  are considered.

\subsection{Search for doubly charged Higgs}

Doubly charged Higgs bosons appear in theories beyond the Standard Model,
like in left-right symmetric models.
In such models the SU(2)$_R$ symmetry is broken by triplet Higgs 
fields which don't conserve baryon and lepton numbers.
The DELPHI and OPAL collaborations have recently searched for pair-produced
doubly charged Higgs bosons. Both collaborations look for 
$\mathrm{H^{++}}$\ra$\mathrm{\tau^+\tau^+}$ decays
while OPAL also searches for \Hp\ra$\mu^+\mu^+$, $e^+,e^+$
decays \cite{hpp}. The data analysed have been taken at 189 $\le$ \sqrts $\le$
209 GeV. A 95\% CL limit on the charged Higgs boson mass of 97.3 \gevc\ for any 
value of the h$_{\tau\tau}$ Yukawa coupling has been obtained by
DELPHI, while OPAL extracts a lower limit of 98.5 \gevc\ for Higgs bosons 
decaying in 100\% of the cases to two leptons of the same sign with 100\% 
branching ratio.
 
\section{Fermiophobic Higgs}

In the minimal Standard Model, the rate of Higgs boson
decays into photons is too small for observation at existing accelerators, but in
other theoretical models, like the 2HDM(I) for certain values of the parameters
the ${\mathrm{h}}^{0}$  couples only to bosons.
The class of ``fermiophobic'' Higgs models includes the more general 
``Bosonic'' Higgs model, 
and Higgs-Triplet models where the particles formed from the triplet fields 
are fermiophobic.
For $m_{\mathrm{h}} <$~80~GeV, the fermiophobic Higgs decays 
primarily into $\gamma\gamma$, while for higher masses decays into
WW$^*$ and \Zo\Zo$^*$ can be observed.\\
The four LEP experiments search for events having two energetic,
isolated photons using data taken at 189 $\le$ \sqrts $\le$ 209 GeV. 
In addition, the \Zo\ decay products are either
classified, or, in the case of \Zo\ra \nunu,
acoplanarity on the photons is required. For the 2000 data,
the ALEPH analysis is ``global'' (the final states are not listed
separately). The  combination of the LEP data
data collected between 88 and 209 GeV in the centre-of-mass 
result in an upper limit on the on the branching ratio B(\h\ra$\gamma\gamma$)
as a function of the Higgs mass~\cite{fermiop}: an observed (expected) lower limit 
on the Higgs boson mass of 109.7 (109.4) \gevc as can be seen in 
Figure~\ref{fig:cpvmssm}. 

\section{Decay mode independent searches}

Searches for neutral scalar bosons \So\ produced in association with the \Zo\ 
have been developed by OPAL in order to search for Higgs bosons with a minimum 
dependence on model assumptions~\cite{dmodeind}.
The analyses are based on studies of the recoil mass spectrum of \Zo\ra\epem\ and 
\mumu\ events and on a search for \So\Zo\ events with \So\ra\epem or photons
and \Zo\ra\nunu.
The limits are sensitive to decays of the \So\ into hadrons, leptons, photons,
invisible particles or if the \So\ lifetime is long enough to escape detection.
The analyses use the full luminosity collected at the \Zo\ peak  at LEP1 and 
662.4 pb$^{-1}$ of LEP2 data collected at 183 $\le$ \sqrts $\le$ 209 GeV.
The results are presented in terms of limits on the scaling factor {\it k}
given by the ratio of the {\it Higgs-strahlung} production cross-section in a given 
model and the SM cross-section.
Values for {\it k} $\ge$ 0.1 are excluded for  \mS $\le$  19 
\gevc\ and
an observed (expected)  lower limit on  \mS\ of 81 (64) \gevc\
is obtained when assuming SM production cross-section. 

\section{Conclusions}

Neutral and charged Higgs bosons have been searched by the LEP collaborations
in any existing and theoretically appealing model that predicts the existence of
Higgs bosons detectable at LEP. Both when searching for
specific models or when using the model independent approaches,
no evidence of a signal has been found and exclusion limits on the parameter
space of the different models have been extracted.

\end{document}